\def\e{{\rm e}}
\def\del{\partial}

\def\abs#1{{\left|{#1}\right|}}
\def\vev#1{\langle #1 \rangle}

\def\del{\partial}
\def\dslash{\del\kern-0.55em\raise 0.14ex\hbox{/}}

\newcommand{\PRD}[3]{Phys. Rev. {\bf D{#1}}, {#2} (19{#3})} 
\newcommand{\PRDM}[3]{Phys. Rev. {\bf D{#1}}, {#2} (20{#3})} 
 
\newcommand{\NPB}[3]{Nucl. Phys. {\bf B{#1}}, {#2} (19{#3})} 
\newcommand{\PLB}[3]{Phys. Lett. {\bf B{#1}}, {#2} (19{#3})}

\newcommand{\PTP}[3]{Prog. Theor. Phys. {\bf {#1}}, {#2} (19{#3})} 
\newcommand{\PTPM}[3]{Prog. Theor. Phys. {\bf {#1}}, {#2} (20{#3})} 
\newcommand{\ANN}[3]{Ann. Phys. (N.Y.) {\bf {#1}}, {#2} (19{#3})}

\newcommand{\hmu}{\hat\mu}

\documentclass[12pt]{article}
\begin{document}
\title{Gauge Symmetry Breaking through the Hosotani Mechanism in 
Softly Broken Supersymmetric QCD}
\author{Kazunori Takenaga \vspace{1cm}$^{}$\thanks {email: 
takenaga@synge.stp.dias.ie}\\ 
\it {School of Theoretical Physics,}\\ 
{\it Dublin Institute for Advanced Studies, 10 Burlington Road, 
Dublin 4,}\\
{\it Ireland}}
\date{} 
\maketitle
\baselineskip=18pt
\vskip 2cm
\begin{abstract}
Gauge symmetry breaking through the Hosotani mechanism (the 
dynamics of nonintegrable phases) in softly 
broken supersymmetric QCD with $N_F^{fd}$ flavors is studied. 
For $N=$ even, there is a single $SU(N)$ symmetric vacuum 
state, while for $N=$ odd, there is a doubly degenerate 
$SU(N)$ symmetric vacuum state in the model. 
We also study generalized supersymmetric 
QCD by adding $N_F^{adj}$ numbers of massless 
adjoint matter. The gauge symmetry breaking pattern
such as $SU(3)\rightarrow SU(2)\times U(1)$ 
is possible for appropriate choices of the matter content and 
values of the supersymmetry breaking parameter. 
The massless state of the adjoint 
Higgs scalar is also discussed in the models.
\end{abstract}
\vskip 2cm
\begin{flushleft}
DIAS-STP-02-04\\
May 2002\\
\end{flushleft}
\begin{center}
\end{center}
\addtolength{\parindent}{2pt}
\newpage
\section{Introduction}
Gauge symmetry breaking through the Hosotani 
mechanism~\cite{hosotani, hosotanib} (the dynamics of nonintegrable 
phases) is one of the 
remarkable phenomena in physics with extra 
dimensions. 
Component gauge fields for compactified 
directions, which are dynamical degrees of freedom and cannot be gauged 
away, can develop vacuum expectation values, and the gauge symmetry
is broken dynamically.
The existence of the zero mode for the component gauge
field is crucial for the mechanism. Quantum effects shift the zero mode
to induce the gauge symmetry breaking, reflecting the topology of the
extra dimension. 
\par
The vacuum expectation values, which are 
nothing but the constant background fields, are also related with the 
eigenvalues (phases) of the Wilson line 
integrals along the compactified direction, and 
the gauge symmetry breaking corresponds to 
the nontrivial Wilson line integral.
One can discuss the gauge symmetry breaking patterns of the theory
by studying the effective potential for the phases~\cite{hosotanib}.
\par
Since the pioneering work by Hosotani~\cite{hosotani}, the 
dynamics of the nonintegrable
phases has been studied in various 
models~\cite{hosotanib}--\cite{inte}, namely, nonsupersymmetric 
gauge models. It has been known that the gauge 
symmetry breaking patterns depend 
on matter content, i.e., the number, the boundary conditions of the 
fields, and the representation under the gauge group of matter. 
\par
In this paper, following the author's works~\cite{takenaga, takenagab}, we 
study the gauge symmetry breaking patterns in  
supersymmetric $SU(N)$ gauge theory with $N_F^{fd}$ numbers of massless
fundamental matter (supersymmetric QCD) defined on $M^3\otimes S^1$.
Here $M^3, S^1$ are three-dimensional Minkowski space-time and a 
circle, respectively. And we also study  
generalized supersymmetric QCD
(supersymmetric QCD with massless adjoint matter). 
\par
The dynamics of the nonintegrable phases determines the vacuum structure 
of the theory. If we, however, introduce the matter multiplets, the 
vacuum expectation values of the squark fields
in the multiplets also become the order parameters for gauge symmetry 
breaking. We assume that the gauge coupling constant $g$ is small and 
ignore $O(g^2)$ contributions to the effective potential. 
In this approximation, there exist flat directions of the potential 
parametrized by the vacuum expectation values of the squark field. 
In order to concentrate on the dynamics of the
nonintegrable phases, we take the trivial ``point'' on the flat 
direction, where all the vacuum expectation values of the squark fields
vanish.
\par
If the theory has supersymmetry, one cannot discuss the dynamical breaking
of gauge symmetry based on perturbation theory because 
the perturbative effective potential for the nonintegrable
phases vanishes due to the supersymmetry. One must break the
supersymmetry in order to obtain nonvanishing effective 
potential\footnote{This is not the case where the gauge charge such 
as the gauged $U(1)_R$ in supergravity models
distinguishes bosons and fermions in a supermultiplet. 
In this case supersymmetry is broken spontaneously by the 
Hosotani mechanism~\cite{riotto}.}. We resort to the Scherk-Schwarz 
mechanism~\cite{ss, fi}, which is a natural candidate to 
break supersymmetry softly in this setup~\cite{takenagac}.
\par
In the softly broken supersymmetric Yang-Mills theory, the $SU(N)$ gauge
symmetry is not broken through the Hosotani mechanism. There are $N$ vacuum
states in the model, and the vacuum has $Z_N$ symmetry. By adding 
$N_F^{fd}$ sets of massless fundamental matter multiplet, the model 
describes the softly broken supersymmetric QCD with $N_F^{fd}$ flavors.
We find that in the case $N=$ even, there is a single $SU(N)$ symmetric 
vacuum state, while in the case $N=$ odd, there 
is a doubly degenerate $SU(N)$ symmetric vacuum state in the model. 
The degenerate two vacua are related to each other by the symmetry
transformations of the effective potential.
Unlike the case of the softly broken supersymmetric Yang-Mills theory,
there is no $Z_2$ symmetry for the degenerate vacuum 
because of the fundamental matter in the model.
The vacuum configurations do not depend on 
the values of $N_F^{fd}$ and the supersymmetry breaking parameter.
\par
We also discuss the mass of the adjoint Higgs scalar. The scalar 
is originally the component gauge field
for the $S^1$ direction and behaves as adjoint Higgs scalar at low energies. 
It acquires mass through the quantum correction in the extra 
dimension, and the mass is
obtained by evaluating the second derivative of the effective potential
at the minimum. 
The adjoint Higgs scalar is always massive in the softly 
broken supersymmetric QCD.
\par
In the generalized supersymmetric QCD, we 
find that the partial gauge symmetry breaking 
such as $SU(2)\times U(1)$, which may be important for grand unified 
theory (GUT) symmetry 
breaking, is possible
for appropriate choices of the matter content and values of
the supersymmetry breaking parameter.
This gauge symmetry breaking pattern is not realized until one
considers both the massless adjoint and fundamental matter multiplets.
We also find the massless state of
the adjoint Higgs scalar within our approximation 
for the aforementioned gauge symmetry breaking pattern in the model.
\par
In the next section we present the effective potentials for the
nonintegrable phases of the models we study in this paper. 
And we determine the gauge symmetry breaking patterns
in the softly broken supersymmetric QCD. The massless adjoint 
Higgs scalar is also 
discussed in the model.
In section $3$ we consider the 
generalized supersymmetric QCD.
We are, especially, interested in the gauge
symmetry breaking pattern such as $SU(3)\rightarrow SU(2)\times U(1)$ and the 
massless state of the adjoint Higgs scalar. 
The final section is devoted to conclusions and discussion.
\par
\section{Supersymmetric QCD with $N_F^{fd}$ flavors}
\subsection{Effective potential for nonintegrable phases}
We present the effective potentials for the nonintegrable phases 
of our models. The effective potential for the 
phase is computed by expanding the fields around the constant 
background gauge field,
\begin{equation}
\vev{A_y}\equiv {1\over{gL}}\vev{\Phi}={1\over {gL}}
{\rm diag}(\theta_1, \theta_2,\cdots, \theta_{N})\qquad {\rm with}\qquad
\sum_{i=1}^N\theta_i=0,
\label{background}
\end{equation} 
and $\theta_i$ is related to the Wilson line integral,
\begin{equation}
W_c\equiv {\cal P}{\rm exp}\left({-ig \oint_{S^1}dy \vev{A_y}}\right)=
{\rm diag}\left(\e^{-i\theta_1}, \e^{-i\theta_2}, \cdots, \e^{-i\theta_N}
\right).
\end{equation}
The residual gauge symmetry is generated by the 
generators of $SU(N)$ commuting with $W_c$~\cite{hosotanib}. 
Following the standard technique given in the 
papers~\cite{hosotani, hosotanib}, 
the effective potential for the softly broken $SU(N)$ supersymmetric 
Yang-Mills theory has been obtained as~\cite{takenaga},
\begin{equation}
V_{SYM}(\theta)={{-2}\over {\pi^2L^4}}
\sum_{n=1}^{\infty}\sum_{i,j=1}^{N}{1\over n^4}
\left(\cos[n(\theta_i-\theta_j)]
-\cos[n(\theta_i-\theta_j-\beta)]\right).
\label{symeff}
\end{equation}
The nontrivial phase $\beta$, which breaks supersymmetry softly, comes from
the boundary condition associated with the $U(1)_R$ symmetry on the gaugino
field~\cite{takenaga, takenagac}.
\par
Let us introduce $N_F^{fd}$ sets of fundamental massless matter 
multiplet denoted by $Q (\bar Q)$ belonging 
to the (anti)fundamental representation
under $SU(N)$. The physical fields in $Q ({\bar Q})$
are quark $q ({\bar q})$ and squark $\phi_q ({\bar\phi}_q)$.
We impose the boundary conditions associated with 
the $U(1)_R$ symmetry on the squark 
fields\footnote{These boundary 
conditions are defined by the assignments of $U(1)_R$ charge
on the fields based on the invariance of the action under the
$U(1)_R$ transformation in the presence of
the mass term $m{\bar Q}Q$. 
The discussion on the effective potential of the 
nonintegrable phases in this paper corresponds to the 
massless limit.}~\cite{takenaga}, $\phi_q ({\bar\phi}_q)(x^{\mu}, y+L)
=\e^{i\beta}\phi_q ({\bar\phi}_q)(x^{\mu}, y)$,
where we have suppressed the flavor index for the squark.
It has been pointed out that the
phase is common to all flavors, so that the 
supersymmetry breaking terms in three dimensions are flavor 
blind~\cite{takenaga, takenagac}.
\par
In order to evaluate the effective potential for the
phases, one needs the mass operators for $Q$ and $\bar Q$, which actually 
give the mass terms for the (s)quarks in three
dimensions after compactifications. 
Since the matter multiplet $Q ({\bar Q})$ belongs to the (anti)fundamental
representation under $SU(N)$ and the squark fields
have the nontrivial phase $\beta$, the mass operator for $\phi_q$ 
and that for $\bar{\phi_q}$ have different 
forms\footnote{This point has 
been overlooked in the previous paper~\cite{takenaga}.}. On the other 
hand, the quark fields have no nontrivial phase, so that 
both $q$ and $\bar q$ give the same mass 
operators\footnote{This is also clear from the fact that $q$ 
and $\overline{\bar q}$ forms a Dirac spinor satisfying the periodic 
boundary condition.}.
\par
One can read the mass operators in the covariant derivatives for
the squark fields,
\begin{equation}
(\del_{\hmu}\phi_q^{\dagger}+ig\phi_q^{\dagger}A_{\hmu})
(\del_{\hmu}\phi_q-ig A_{\hmu}\phi_q), \qquad
(\del_{\hmu}{\bar\phi}_q+ig{\bar\phi}_qA_{\hmu})
(\del_{\hmu}{\bar\phi}_q^{\dagger}-ig A_{\hmu}{\bar\phi}_q^{\dagger}).
\end{equation}
They are obtained as 
\begin{eqnarray}
\left(D_3^{\phi_q}\right)^2&=&-\sum_{n=-\infty}^{\infty}
\sum_{i=1}^{N}\left({{2\pi}\over{L}}\right)^2
\left(n-{{\theta_i-\beta}\over {2\pi}}\right)^2
\qquad {\rm for}\qquad \phi_q, \label{massopa}\\
\left(D_3^{\bar{\phi}_q}\right)^2&=&
-\sum_{n=-\infty}^{\infty}\sum_{i=1}^{N}
\left({{2\pi}\over{L}}\right)^2\left(n-{{-\theta_i-\beta}\over {2\pi}}
\right)^2
\qquad {\rm for}\qquad {\bar\phi_q}. 
\label{massopb}
\end{eqnarray}  
Here $n$ stands for the Kaluza-Klein mode for the $S^1$ direction.
That the prescription $\theta_i\rightarrow -\theta_i$
in Eq. (\ref{massopa}) gives Eq. (\ref{massopb}) shows
the field $\bar{\phi_q}$ belongs to the antifundamental 
representation under $SU(N)$. We see that  
$\phi_q$ and ${\bar\phi}_q$ contribute to 
the effective potential 
in a different manner\footnote{The gauge group $SU(2)$ is an exceptional
case as we will see in the section $3$.}.
\par
Following again the standard prescription, we obtain the effective potential
for the phases coming from the fundamental massless matter multiplets,
\begin{eqnarray}
V_{matter}^{fd}(\theta)&=&
{{2N_F^{fd}}\over{\pi^2L^4}}\sum_{n=1}^{\infty}
\sum_{i=1}^{N}{1\over n^4}\left((\cos(n\theta_i)-\cos[n(\theta_i-\beta)])
+(\cos(n\theta_i)-\cos[n(\theta_i+\beta)])\right)\nonumber\\
&=& 
{{2N_F^{fd}}\over{\pi^2L^4}}\sum_{n=1}^{\infty}
\sum_{i=1}^{N}{1\over n^4}(2\cos(n\theta_i)-\cos[n(\theta_i-\beta)]
-\cos[n(\theta_i+\beta)])
\label{fdeff}
\end{eqnarray}
where the first term in Eq. (\ref{fdeff}) arises from the 
quarks $q, {\bar q}$, and the
second and third terms come from $\phi_q$ and ${\bar\phi}_q$, respectively.
By putting Eqs. (\ref{symeff}) and (\ref{fdeff}) 
together, we obtain the effective potential for 
the softly broken supersymmetric QCD with $N_F^{fd}$
numbers of the massless fundamental matter,
\begin{eqnarray}   
V_{SQCD}(\theta)&=&V_{SYM}(\theta)+V_{matter}^{fd}(\theta)\nonumber\\
&=&{{-2}\over {\pi^2L^4}}
\sum_{n=1}^{\infty}\sum_{i,j=1}^{N}{1\over n^4}(\cos[n(\theta_i-\theta_j)]
-\cos[n(\theta_i-\theta_j-\beta)])\nonumber\\
&+&{{2N_F^{fd}}\over{\pi^2L^4}}\sum_{n=1}^{\infty}
\sum_{i=1}^{N}{1\over n^4}(2\cos(n\theta_i)-\cos[n(\theta_i-\beta)]
-\cos[n(\theta_i+\beta)]).
\label{sqcdeff}
\end{eqnarray}
\par
As a general remark, the phase $\theta_i$ gives no physical 
effects at least classically, but the effect is essential at 
the quantum level. It should be emphasized that these effective 
potentials (\ref{symeff}), (\ref{fdeff})
arise from taking into account the quantum correction 
in the extra dimension.
\subsection{Gauge symmetry breaking via the Hosotani mechanism}
We discuss the gauge symmetry breaking through
the Hosotani mechanism based on the obtained effective potentials in the
previous subsection. Before doing this, let us mention 
the vacuum structure of the model, which is peculiar to softly broken
supersymmetric gauge theories. 
\par
Strictly speaking, the dynamics of nonintegrable phases
itself does not give complete information on the vacuum structure of 
softly broken supersymmetric gauge theories.
This is because, as noted in the Introduction, the vacuum 
expectation values of the squark fields $\vev{\phi_q}, \vev{{\bar\phi}_q} 
\in \rm {\bf C}$ are also the order parameters for gauge symmetry 
breaking. If one wishes to study the entire vacuum structure, one should 
take into account the order 
parameters in addition to the nonintegrable phases.
This means that one has to include the
tree-level potential and one-loop corrections to the vacuum
expectation values of the squark fields as well. 
\par
The tree-level potential, which arises from the covariant 
derivative and the quartic couplings for the squark field, is 
given by\footnote{Since the tree-level potential in the model 
is not the Higgs type potential, we do not expect the phase structures 
depending on the size of $S^1$ such as the ones studied 
in Ref.~\cite{sakamoto}.}.
\begin{eqnarray}
V_{tree}&=&g^2\left(\vev{\phi_q^{\dagger}}\vev{A_y}^2\vev{\phi_q}+
\vev{{\bar\phi}_q}\vev{A_y}^2\vev{{\bar\phi}_q^{\dagger}}\right)+g^2
\left(\vev{\phi_q^{\dagger}}T^a\vev{\phi_q}-\vev{{\bar\phi_q}}T^a
\vev{{\bar\phi_q}^{\dagger}}\right)^2
\nonumber\\
&=&{1\over L^2}
\sum_{i=1}^{N}\theta_i^2\left(\abs{\vev{\phi_{qi}}}^2
+\abs{\vev{{\bar\phi}^{i}_q}}^2
\right)
+g^2\left(\vev{\phi_q^{\dagger}}T^a\vev{\phi_q}
-\vev{{\bar\phi_q}}T^a\vev{{\bar\phi_q}^{\dagger}}
\right)^2,
\end{eqnarray}
where we have used Eq. (\ref{background}) 
and $T^a (a=1,\cdots, N^2-1)$ stands for the 
generator of $SU(N)$. Let us note that the interactions between
$\vev{\phi_q}, \vev{{\bar\phi}_q}$ and $\theta_i$ are $O(1)$, while 
the self-interactions among the squarks
are of order $g^2$. And the one-loop 
correction to the vacuum expectation values of the 
squark fields, which is not written explicitly, is also of order $g^2$.
\par
If the gauge coupling $g$ is very small, then, one may ignore the 
$O(g^2)$ terms, so that the term which does not have the gauge coupling 
dependence becomes a dominant contribution to the vacuum structure 
of the theory.
In this approximation, the total effective potential is given by
\begin{equation}
V(\theta, \vev{\phi_q}, \vev{{\bar\phi}_q})=
{1\over L^2}
\sum_{i=1}^{N}\theta_i^2\left(\abs{\vev{\phi_{qi}}}^2
+\abs{\vev{{\bar\phi}^{i}_q}}^2\right)+V_{SQCD}(\theta),
\label{totaleff}
\end{equation}
where $V_{SQCD}(\theta)$ is given by Eq. (\ref{sqcdeff}). 
The relevant interaction to generate the 
effective potential (\ref{totaleff}) is only the
gauge interaction, which is $O(1)$. That is why the total effective
potential does not have the dependence on the gauge coupling.
\par
The first term in Eq. (\ref{totaleff}), which stands for the 
tree-level potential, is positive semi-definite.
The configuration that minimizes it
is given by $\vev{\phi_{qi}}=\vev{{\bar\phi}_q^i}=0$ for nonzero values 
of $\theta_i (i=1,\cdots, N)$. In fact, as we will see soon, the 
nonzero values of $\theta_i$ are the case where  
the absolute minima of $V_{SQCD}(\theta)$ is realized.
As a result, the tree-level potential 
does not affect the vacuum structure
of the model in this approximation. Therefore, the vacuum structure is 
determined by the dynamics of the nonintegrable phases alone in this model.
\par
Let us now consider the effective potential $V_{SQCD}(\theta)$ in 
order to study the dynamics of the nonintegrable phases, i.e., gauge
symmetry breaking through the Hosotani mechanism. 
Our strategy to find the vacuum configuration of
the potential is to minimize $V_{SYM}(\theta)$ 
and $V_{matter}(\theta)$ separately, and we take 
the common configuration for both of them, which actually gives the 
absolute minima of the potential $V_{SQCD}(\theta)$. It has been
studied~\cite{takenagab} that the absolute minima
of $V_{SYM}(\theta)$ is located at
\begin{equation}
\theta_i(i=1,\cdots, N)={{2\pi}\over{N}}m, \qquad m=0, \cdots, N-1.
\label{symsol}
\end{equation}
The Wilson line integral just corresponds to an element of the 
center of $SU(N)$, so that the gauge symmetry is not broken.
\par
It is important to note here that there are $N$ vacuum states
corresponding to the values of $m$.
The $N$ vacua are physically equivalent
because, for example, the mass spectra on the vacua are exactly the 
same as each other. The fields $A_{\mu}, \lambda$ remain 
massless on the vacuum configuration (\ref{symsol}).
The vacuum has $Z_N$ symmetry. 
A way of looking at the $Z_N$ symmetry is to 
consider the gauge transformation (regular, nonperiodic) defined by
\begin{equation}
U^{(m)}(y)={\rm exp}{{2\pi i y}\over L}
\left(\begin{array}{cccc}
{m\over N}& & & \\
&{m\over N} & &\\
& & \ddots & \\
& & &-{{(N-1)m}\over N}\end{array}\right).
\label{gaugetrf}
\end{equation}
This transformation does not change the boundary conditions of the fields 
$A_{\hmu}, \lambda$ because they belong to the adjoint representation
under $SU(N)$. It is easy to see that the $N$ vacuum states are 
related to each other by this transformation.
\par
Let us next consider the potential $V_{matter}^{fd}(\theta)$ given 
by Eq. (\ref{fdeff}) and find 
the configuration that minimizes it. 
This is interesting in its own light because, as we will 
see later, this potential corresponds to the case of 
the generalized supersymmetric QCD with $N_{F}^{adj}=1$. 
The potential is recast as
\begin{equation}
V_{matter}^{fd}(\theta)
={{4N_F^{fd}}\over {\pi^2 L^4}}\sum_{n=1}^{\infty}
\sum_{i=1}^{N}{1\over n^4}[1-\cos(n\beta)]\cos(n\theta_i).
\end{equation}
We see that the nontrivial phase $\beta$ does not affect the
location of the absolute minima of the potential. In finding the
minimum, let us note that the potential is invariant 
under\footnote{The potential is also invariant 
under $\beta\rightarrow \beta + 2\pi i k, k\in {\bf Z}$. 
This corresponds to $\lambda\rightarrow \e^{2\pi i k}\lambda$.}
\begin{equation}
\beta\rightarrow 2\pi -\beta.
\label{inva}
\end{equation}
This invariance means that the potential is symmetric 
under the reflection with respect to $\beta=\pi$
for fixed $\theta_i$. 
The region given by $0<\beta \leq \pi$ is enough to
study the potential. Moreover, the potential also possesses
the invariance under
\begin{equation}
\theta_i\rightarrow 2\pi-\theta_i,\qquad i=1, \cdots, N.
\label{invb}
\end{equation}
The maximal symmetry of $V_{matter}^{fd}(\theta)$ is given by the 
transformations with Eqs. (\ref{inva}) and (\ref{invb}).
\par
Taking into account Eqs. (\ref{inva}) and (\ref{invb}), we see that the 
region given by $\theta_i-\beta \geq 0$ is enough to study the
potential. Thanks to this, one does not need the classification
depending on the sign of $\theta_i-\beta$ when one uses the formula
\begin{equation}
\sum_{n=1}^{\infty}{1\over{n^4}}\cos(nx)={{-1}\over{48}}x^2(x-2\pi)^2+
{{\pi^4}\over{90}} \qquad (0\leq x\leq 2\pi).
\label{forma}
\end{equation}
Noting an expression obtained by applying the formula (\ref{forma}),  
\begin{equation}
\sum_{n=1}^{\infty}{1\over n^4}(2\cos(n\theta)-\cos[n(\theta-\beta)]-
\cos[n(\theta+\beta)])
={{\beta^2}\over {24}}(6\theta^2-12\pi\theta + \beta^2 +4\pi^2),
\label{formb}
\end{equation}
we have
\begin{equation}
V_{matter}^{fd}={{2N_F^{fd}}\over {\pi^2 L^4}}{\beta^2\over {24}}
\Biggl(\sum_{i=1}^{N-1}\Bigl(6\theta_i^2 -12\pi\theta_i +\beta^2+4\pi^2\Bigr)
+6\Bigl(\sum_{i=1}^{N-1}\theta_i\Bigr)^2-12\pi\sum_{i=1}^{N-1}\theta_i
+\beta^2+4\pi^2\Biggr).
\end{equation}
The extremum condition $\del V_{matter}/\del\theta_k (k=1,\cdots, N-1)=0$ 
yields
\begin{equation}
\theta_k+(\theta_1+\cdots +\theta_{N-1})=0~~({\rm mod}~~2\pi), 
\qquad k=1,\cdots, N-1.
\label{equation}
\end{equation}
The solution to Eq. (\ref{equation}) is obtained as 
$\theta_k=2\pi q/N~(q=0, \cdots, N-1)$.
Since $\theta_N=-\sum_{k=1}^{N-1}\theta_k=2\pi q/N$, we finally have
$\theta_i(i=1,\cdots, N)=2\pi q/N$.
\par
Unlike the case of the softly broken supersymmetric 
Yang-Mills theory, the effective potential has different energies
for different values of $q$ in the present case. 
The minimum of the function (\ref{formb})
is achieved at $\theta=\pi$. If all the $\theta_i$'s can 
take this value, the potential $V_{matter}^{fd}(\theta)$ is 
obviously minimized at $\theta_i=\pi~(i=1,\cdots, N)$. 
In fact, this is the case when $N=$ even and corresponds to $q_{even}=N/2$. 
For $N=$ odd, the value which is as close as 
possible to $\pi$ gives the lowest energy of 
the potential. It is given 
by $q_{odd}^{(1)}=(N-1)/2$, i.e., $\theta_i^{(1)}=(N-1)\pi/N$. 
The potential is invariant under Eq. (\ref{invb}), so that the configuration
with $q_{odd}^{(2)}=(N+1)/2$ corresponding to 
$\theta_i^{(2)}=(N+1)\pi/N (=2\pi-\theta_i^{(1)})$ gives the same energy 
as that for $q_{odd}^{(1)}=(N-1)/2$ and also becomes a vacuum 
configuration\footnote{Note that the physical
region of $\theta_i (i=1,\cdots, N)$ is $0\leq \theta_i \leq 2\pi$.}.
\par
The two vacuum configurations $\theta_i^{(1)}, \theta_i^{(2)}$ are
not distinct. In order to see this, let us
consider the mass spectra for $\phi_q$ on the 
vacua $\theta_i^{(1)}, \theta_i^{(2)}$. They are given by
$(n-(\theta_i^{(1)}-\beta)/2\pi)^2$ 
and $(n-(\theta_i^{(2)}-\beta)/2\pi)^2$ from Eq. (\ref{massopa}).
The former is reduced to the latter by the transformations
with Eqs. (\ref{inva}) and (\ref{invb}) and vice versa. 
Since they are the symmetry transformation of the effective potential, both 
of the mass spectra are physically identical to each other.
\par
The vacuum configuration for the case $N=$ odd is a doubly degenerate.
There is, however, no $Z_2$ symmetry for the vacuum configurations in
the present case because the model contains the massless matter 
multiplet belonging to the (anti)fundamental
representation under $SU(N)$. The gauge transformation 
with Eq. (\ref{gaugetrf})
changes the boundary condition of the field in the multiplet. 
In fact, we see that 
\begin{equation}
\phi_q^{\prime}(y+L)=\e^{i(\beta+{{2\pi}\over N})}\phi_q^{\prime}(y),
\end{equation}
where $\phi_q^{\prime}=U^{(m=1)}(y)\phi_q$.
\par
We have obtained the vacuum configuration which 
minimizes $V_{matter}^{fd}(\theta)$ as
\begin{equation}
\theta_i (i=1,\cdots, N)
=\left\{\begin{array}{lcl} \pi & & \cdots N={\rm even}, \\[0.2cm]
{{N-1}\over N}\pi, & ({\rm or}~~{{N+1}\over N}\pi) &\cdots N={\rm odd}.
\end{array}\right.
\label{fdsol}
\end{equation}
As we have noticed before, they do not depend on $N_F^{fd}$ and the 
supersymmetry breaking parameter $\beta$ by 
the Scherk-Schwarz mechanism. 
The vacuum configurations 
respect the $SU(N)$ gauge symmetry and are the parts of the center
of $SU(N)$. 
\par
We are ready to find the common configuration
between Eqs. (\ref{symsol}) and (\ref{fdsol}), which gives 
the absolute minima of the effective potential (\ref{sqcdeff}).
It is given by Eq. (\ref{fdsol}) obviously.
We conclude that for $N=$ even, there is a single vacuum state, while 
for $N=$ odd, there is a doubly degenerate vacuum state 
in the softly broken supersymmetric QCD with $N_F^{fd}$ flavors. 
\par
Here we confirm the discussion on the tree-level potential 
at the beginning of this subsection. As we have studied above, the 
configuration that minimizes the effective 
potential (\ref{sqcdeff}) is given by the nonzero 
values of $\theta_i(i=1,\cdots,N)$, so that 
only the vanishing vacuum expectation values of the
squark fields minimize the total potential (\ref{totaleff}).
\par
Let us now study the mass of the adjoint Higgs scalar. The scalar
is originally the component gauge field for the $S^1$ direction 
and behaves as an adjoint Higgs scalar at low energies.
It acquires mass through the quantum correction in the extra dimension.
The mass is obtained by the second derivative 
of the effective potential (\ref{sqcdeff}) at the minimum,
\begin{equation}
{{\del^2V_{SQCD}}\over{\del\theta_i\del\theta_j}}
={{C_H^{SQCD}}\over{\pi^2 L^4}}M_{ij},\qquad C_H^{SQCD}\equiv 
\beta^2\left(N+N_F^{fd}\right),
\end{equation}
where the matrix $M_{ij}$ is given by
\begin{equation}
M_{ij}\equiv \left(\begin{array}{ccccc}
2&1&\cdots&\cdots& 1\\
1&2& & &\vdots\\
\vdots& &\ddots & &\vdots\\
\vdots& & & \ddots&\vdots\\
1&\cdots&\cdots&\cdots&2\\
\end{array}\right).
\label{matrix}
\end{equation}
All the (off-)diagonal elements of the matrix are $2(1)$.
As studied in Ref.~\cite{takenagab}, this matrix 
is easily diagonalized, and the mass is obtained as
\begin{equation}
m_{\Phi}^2={{g^2C_H^{SQCD}}\over{\pi^2 L^2}}{N\over 2}.
\end{equation}
The mass of the adjoint Higgs scalar is $SU(N)$ invariant, reflecting 
the $SU(N)$-symmetric vacuum configuration of the model. 
It is easy to see that there is no
possibility of having $C_H^{SQCD}=0$, so that the adjoint Higgs 
scalar is always massive and cannot be massless.
\section{Supersymmetric QCD with massless adjoint matter}
In this section we proceed to study the generalized version of 
supersymmetric QCD by introducing $N_F^{adj}$ numbers of  
massless adjoint matter multiplet.
Let us first discuss the tree-level potential 
within our approximation in this model. 
\par
If we add the massless adjoint matter, the 
tree-level potential becomes, ignoring the $O(g^2)$ terms 
and the flavor index\footnote{We have ignored the terms coming from 
the trilinear coupling of the chiral superfields, ${\bar Q}Q^{adj}Q$, by
assuming that the coupling is of order $g$, hence $O(g^2)$ in the
potential.}.  
\begin{equation}
V_{tree}=
{1\over L^2}
\sum_{i=1}^{N}\theta_i^2\left(\abs{\vev{\phi_{qi}}}^2
+\abs{\vev{{\bar\phi}_q^{i}}}^2\right)
+{{2}\over L^2}{\rm tr}\abs{\Bigl[\vev{\Phi},~\vev{\phi_q^{adj}}\Bigr]}^2.
\end{equation}
The second term comes from the covariant derivative of the
squark field in the adjoint representation under $SU(N)$. 
The total effective potential is, then, given by 
\begin{equation}
V_{total}={1\over L^2}
\sum_{i=1}^{N}\theta_i^2\left(\abs{\vev{\phi_{qi}}}^2
+\abs{\vev{{\bar\phi}_q^{i}}}^2\right)
+{{2}\over L^2}{\rm tr}\abs{\Bigl[\vev{\Phi},~\vev{\phi_q^{adj}}\Bigr]}^2
+V_{GSQCD}(\theta).
\label{totalneweff}
\end{equation}
$V_{GSQCD}(\theta)$ is given by
\begin{eqnarray}
V_{GSQCD}(\theta)&\equiv &V_{SQCD}(\theta)+V_{matter}^{adj}(\theta)
\nonumber\\
&=&{{2N_F^{adj}-2}\over {\pi^2L^4}}
\sum_{n=1}^{\infty}\sum_{i,j=1}^{N}{1\over n^4}(\cos[n(\theta_i-\theta_j)]
-\cos[n(\theta_i-\theta_j-\beta)])\nonumber\\
&+&{{2N_F^{fd}}\over{\pi^2L^4}}\sum_{n=1}^{\infty}
\sum_{i=1}^{N}{1\over n^4}(2\cos(n\theta_i)-\cos[n(\theta_i-\beta)]
-\cos[n(\theta_i+\beta)]),
\label{gsqcdeff}
\end{eqnarray}
where the first line in Eq. (\ref{gsqcdeff}) stands 
for the contributions from the supersymmetric 
Yang-Mills theory and $N_F^{adj}$ numbers of the 
massless adjoint matter~\cite{takenagab}.
\par
Let us note that one cannot rotate $\vev{\phi_q^{adj}}$ into a 
diagonal form by utilizing the $SU(N)$ degrees of freedom
because we have already used them to parametrize $\vev{A_y}$ as 
the diagonal form given by Eq. (\ref{background}). 
The first and second terms in Eq. (\ref{totalneweff}) are positive 
semi-definite. In order to minimize the second term 
in Eq. (\ref{totalneweff}), $\vev{\phi_q^{adj}}$ have only 
a diagonal form. Then, it commutes with 
$\vev{\Phi}$ for any values of $\theta_i$ and yields the vanishing 
second term $[\vev{\Phi},~\vev{\phi_q^{adj}}]=0$. 
Therefore, $\vev{\phi_q^{adj}}$ is undetermined in 
this approximation and parametrizes the flat direction of the potential.
\par
In addition to $\vev{\phi_q^{adj}}$, the vacuum 
expectation values of $\phi_q$ and ${\bar\phi}_q$ can also 
parametrize the flat direction of the potential. 
If all the $\theta_i$'s take nonzero values, 
$\vev{\phi_q}=\vev{{\bar\phi}_q}=0$ gives the 
vanishing first term in Eq. (\ref{totalneweff}). 
In this case, there is no flat direction
of the potential parametrized by $\vev{\phi_q}$ and $\vev{{\bar\phi}_q}$. 
This was the situation in the softly 
broken supersymmetric QCD. 
If some of $\theta_i$'s, however, take the values of 
zero, say, $\theta_k\neq 0 (k=1,\cdots, l< N-1)$, the corresponding 
$\vev{\phi_{qk}}$ and $\vev{{\bar\phi}_q^{k}}$ can take 
arbitrary values in keeping the vanishing first term and parametrize
the flat direction of the potential. In our approximation
ignoring the $O(g^2)$ terms, the effective potential has the 
flat direction in general.
\par
In this paper, we are interested in the dynamics of the nonintegrable 
phases, or one can say that 
we study the gauge symmetry breaking in this
model at the trivial ``point,'' where all the vacuum expectation 
values of the squark 
fields $\phi_q, {\bar\phi}_q, \phi_q^{adj}$ vanish.  We  
ignore the tree-level potential, first and second terms in 
Eq. (\ref{totalneweff}) and  
focus on the effective potential $V_{GSQCD}(\theta)$ only.
\par
Here we notice that the effective potential $V_{GSQCD}(\theta)$ is 
reduced to $V_{matter}^{fd}(\theta)$ for $N_F^{adj}=1$. 
The contributions from the vector multiplet 
$(A_{\hmu}, \lambda)$ and the massless adjoint 
multiplet $(q^{adj},\phi_q^{adj})$ to the constant 
background (\ref{background})
cancel each other. This is because in four dimensions the two 
massless multiplets form ${\cal N}=2$ supersymmetry 
to have the $SU(2)_R$ symmetry, so that we still have 
${\cal N}=1$ supersymmetry for the two multiplets even though
we imposed the boundary condition
associated with the $U(1)_R$ symmetry~\cite{takenaga}. 
As we have already studied in the previous 
section, the vacuum configuration for this special case is 
given by Eq. (\ref{fdsol}) from 
the potential $V_{matter}^{fd}(\theta)$ alone. The $SU(N)$ gauge 
symmetry is not broken for any 
values of $N_F^{fd}$ and $\beta$. In order to avoid the cancellation, one 
needs to impose the boundary condition
associated with the $SU(2)_R$ symmetry in addition to $U(1)_R$.
\subsection{$SU(2)$ case}
The effective potential (\ref{gsqcdeff}) seems to have a simple form.
It is, however, hard to study the vacuum configuration of the
potential fully analytically.
As we will show in the next subsection, the location 
of the minima of the potential changes 
according to the values of the phase $\beta$. 
The only exceptional case is the $SU(2)$ gauge group.
The effective potential for the case of $SU(2)$ becomes
\begin{eqnarray}
V_{GQSCD}(\theta)&=&
{{2N_F^{adj}-2}\over {\pi^2 L^4}}\sum_{n=1}^{\infty}{1\over n^4}
\biggl(2(1-\cos(n\beta))\nonumber\\
&+&2\cos(2n\theta)-\cos[n(2\theta-\beta)]
-\cos[n(2\theta+\beta)]\biggr)\nonumber\\
&+&{{2\times (2N_F^{fd})}\over{\pi^2 L^4}}\sum_{n=1}^{\infty}{1\over n^4}
(2\cos(n\theta)-\cos[n(\theta-\beta)]-\cos[n(\theta+\beta)]).
\label{su2}
\end{eqnarray}
Let us note that the contributions from $\phi_q$ and ${\bar\phi}_q$ 
to the potential (\ref{su2}) have the same forms. This is because
${\bf 2}$ and ${\bar {\bf 2}}$ of $SU(2)$ are equivalent.
The $SU(2)$ gauge group is special in this sense.
\par
The potential (\ref{su2}) happens to be invariant under 
Eqs. (\ref{inva}) and (\ref{invb}), so that
the region given by $\theta-\beta \geq 0$ is enough to study the 
potential, and we can apply the formula (\ref{forma}) to the effective
potential. We obtain that
\begin{eqnarray}
V_{GQSCD}(\theta)&=&
{{2N_F^{adj}-2}\over {\pi^2 L^4}}{{\beta^2}\over {24}}
\left((\beta-2\pi)^2+24\theta^2
-24\pi\theta+\beta^2+4\pi^2\right)\nonumber\\
&+&{{2\times (2N_F^{fd})}\over{\pi^2 L^4}}{{\beta^2}\over{24}}
\left(6\theta^2-12\pi\theta+\beta^2+4\pi^2\right).
\end{eqnarray}
By solving the extremum condition $\del V_{QSCD}/\del\theta=0$, we have
\begin{equation}
\theta^{(1)}={{N_F^{fd}+N_F^{adj}-1}\over{N_F^{fd}+2(N_F^{adj}-1)}}\pi.
\label{su2sol}
\end{equation}
The other solution, which is 
obtained by taking into account the invariance of the potential 
under Eq. (\ref{invb}), 
\begin{equation}
\theta^{(2)}=2\pi-\theta^{(1)}
={{N_F^{fd}+3(N_F^{adj}-1)}\over{N_F^{fd}+2(N_F^{adj}-1)}}\pi
\end{equation}
is not distinct from the solution $\theta^{(1)}$. 
The squark mass spectra on the solutions are identical to each other
due to Eq. (\ref{inva}) [and/or Eq. (\ref{invb})].
There is a doubly degenerate vacuum state. The vacuum configuration breaks 
the $SU(2)$ gauge symmetry to $U(1)$ spontaneously.
\par
The second derivative of the effective potential at the minimum 
gives the mass of the adjoint Higgs scalar as we have stated
in the section $3$. We find that
\begin{equation}
m^2_{\Phi}\equiv (gL)^2{{\del^2 V_{GSQCD}}\over{\del\theta^2}}
={{2g^2\beta^2}\over{\pi^2 L^2}}\Bigl(2(N_F^{adj}-1)+N_F^{fd}\Bigr).
\end{equation} 
No massless state of the adjoint Higgs scalar appears except 
for $(N_F^{adj},N_F^{fd})=(1,0)$, whose flavor number corresponds 
to the aforementioned ${\cal N}=2$ supersymmetry in four dimensions.
\par
\subsection{$SU(3)$ case}
Let us next consider the $SU(3)$ gauge group. 
Even in this case, we find interesting physics such
as the partial gauge symmetry breaking and massless adjoint 
Higgs scalar, which is never observed in the models 
studied in Ref.~\cite{takenagab} and the previous section. 
\par
In order to see that the vacuum configuration changes
according to the values of $\beta$ by the Scherk-Schwarz 
mechanism, we first assume that $\beta$ is
very small, but nonzero. After finding the vacuum configuration 
for the small values of $\beta$, we next study the vacuum configuration 
for $\beta=\pi$. The potential (\ref{gsqcdeff}) for the case 
of $SU(3)$ is still invariant under Eq. (\ref{inva}), so 
that $0<\beta\leq \pi$ is relevant. Then, we compare 
the configurations for the two cases.
\par
We may apply the formula (\ref{forma}) to the 
potential (\ref{gsqcdeff}) for the small 
values of $\beta$. We obtain that
\begin{eqnarray}
V_{GQSCD}(\theta)&=&
{{2N_F^{adj}-2}\over{\pi^2 L^4}}\beta^2
\Biggl[{N\over{48}}(\beta-2\pi)^2+{{N(N-1)}\over{48}}(\beta^2+4\pi^2
)\nonumber\\
&+&{{N}\over{2}}\left(\sum_{i=1}^{N-1}\theta_i^2+\sum_{1\leq i<j\leq N-1}
\theta_i\theta_j\right)-\pi\sum_{i=1}^{N-1}(N-i)\theta_i\Biggr]\nonumber\\
&+&{{4N_F^{fd}}\over{\pi^2 L^4}}{\beta^2\over {48}}
\Biggl[12\left(\sum_{i=1}^{N-1}\theta_i^2+\Bigl(\sum_{i=1}^{N-1}\theta_i
\Bigr)^2\right)
-48\pi\sum_{i=1}^{N-1}\theta_i+N(\beta^2+4\pi^2)\Biggr],
\label{expand}
\end{eqnarray}
where we have used the result obtained in Ref. \cite{takenagab} for the
first line in Eq. (\ref{gsqcdeff}).
The extremum condition $\del V_{GSQCD}/\del\theta_k (k=1, \cdots, N-1)=0$ 
yields
\begin{equation}
{1\over{2\pi}}\left(N(N_F^{adj}-1)+N_F^{fd}\right)
\left(\theta_k+(\theta_1+\cdots +\theta_{N-1})\right)
=N_F^{fd}+(N_F^{adj}-1)(N-k).
\label{gsqcdsol}
\end{equation}
This is written in the form, denoting $d\equiv N(N_F^{adj}-1)+N_F^{fd}$, 
\begin{equation}
{d\over{2\pi}}\left(\begin{array}{ccccc}
2&1&\cdots&\cdots& 1\\
1&2& & &\vdots\\
\vdots& &\ddots & &\vdots\\
\vdots& & & \ddots&\vdots\\
1&\cdots&\cdots&\cdots&2\\
\end{array}\right)\left(\begin{array}{c}
\theta_1\\
\theta_2\\
\theta_3\\
\vdots\\
\theta_{N-2}\\
\theta_{N-1}
\end{array}\right)
=N_F^{fd}
\left(\begin{array}{c}
1\\
1\\
1\\
\vdots\\
1\\
1\end{array}\right)
+(N_F^{adj}-1)
\left(\begin{array}{c}
N-1\\
N-2\\
N-3\\
\vdots\\
2\\
1\end{array}\right),
\label{matrixform}
\end{equation}
where the matrix on the left-hand side in Eq. (\ref{matrixform}) 
is the same as the one in Eq. (\ref{matrix}).
The inverse of the matrix is given by
\begin{equation}
{1\over N}\left(\begin{array}{ccccc}
N-1&-1&\cdots&\cdots& -1\\
-1&N-1& & &\vdots\\
\vdots& &\ddots & &\vdots\\
\vdots& & & \ddots&\vdots\\
-1&\cdots&\cdots&\cdots&N-1\\
\end{array}\right).
\label{inverse}
\end{equation}
All the (off-)diagonal elements of the matrix 
are $N-1(-1)$. The solution to Eq. (\ref{gsqcdsol}) is then found to be 
\begin{equation}
\theta_k={{N_F^{fd}}\over d}{{2\pi}\over N}+{{(N_F^{adj}-1)}\over d}
\pi\Bigl(N-(2k-1)\Bigr),\qquad k=1,\cdots, N-1
\label{gsqcdsolb}
\end{equation}
with
\begin{equation}
\theta_N=-\sum_{k=1}^{N-1}\theta_k={{-2\pi}\over d}{{N-1}\over N}
\left(N_F^{fd}+{N\over 2}(N_F^{adj}-1)\right).
\end{equation}
\par
These solutions become 
\begin{equation}
(\theta_1,\theta_2)=\Biggl({2\over 3}\pi,~~
{{N_F^{fd}}\over{3(N_f^{adj}-1)+N_F^{fd}}}{{2\pi}\over 3}\Biggr).
\label{solsu3}
\end{equation}
for the $SU(3)$ gauge group, which is of our interest.
Except for the case of $N_F^{adj}=1$, the configuration
breaks $SU(3)$ to $U(1)\times U(1)$. Therefore, for the small values of 
$\beta$, the gauge symmetry is maximally broken, which still holds for 
the $SU(N)$ gauge group.
As an example, the solutions for certain values of $N_F^{adj}$
and $N_F^{fd}$ are given by
\begin{eqnarray}
(\theta_1, \theta_2)&=&
\left({2\over 3}\pi, {1\over 6}\pi\right)\cdots (N_F^{adj},N_F^{fd})=(2, 1),
\nonumber\\
&=&\left({2\over 3}\pi, {4\over {15}}\pi\right)
\cdots (N_F^{adj},N_F^{fd})=(2, 2),\nonumber\\
&=&\left({2\over 3}\pi, {1\over 3}\pi\right)
\cdots (N_F^{adj},N_F^{fd})=(2, 3).
\label{example}
\end{eqnarray}
\par
Let us next study the vacuum configuration at $\beta=\pi$. 
The possible gauge symmetry breaking 
patterns are\footnote{We have confirmed that the configurations given 
by $(\theta_1, \theta_2)=(0, 0), (\pi/3, \pi/3) $ do not alter 
our discussions.}.
\begin{equation}
SU(3)\rightarrow \left\{\begin{array}{lcl}
SU(3) & \cdots& (\theta_1, \theta_2)=
({2\over 3}\pi, {2\over 3}\pi),\\[0.2cm]
SU(2) \times U(1)& \cdots & (\theta_1, \theta_2)=(\pi, 0) 
+{\rm permutations},\\[0.2cm]
U(1)\times U(1) & \cdots &(\theta_1, \theta_2)
=({2\over 3}\pi, 0)+{\rm permutations}. 
\end{array}\right.
\label{exampleb}
\end{equation}
By studying the determinant of the Hessian,
\begin{equation}
H_{ij}\equiv {{\del^2V_{GSQCD}}\over{\del\theta_i\del\theta_j}}
\bigg|_{\beta=\pi}
\label{hessiana}
\end{equation}
and comparing the potential energy for the given gauge symmetry 
breaking pattern (\ref{exampleb}),
we know the position and stability of the global minima of the
effective potential. And at the same time, as we will see later, the matrix
gives the information on the mass of the adjoint Higgs scalar at $\beta=\pi$.
Depending on the numbers of flavors $N_F^{adj}, N_F^{fd}$, the gauge
symmetry breaking patterns are different.
We obtain\footnote{The configuration for 
the region $3(N_F^{adj}-1)/7 \leq N_F^{fd} < N_F^{adj}-1$ is not 
given by $(\theta_1, \theta_2)=(\pi, 0)$, but is close to it and respects
$U(1)\times U(1)$ symmetry.}
\begin{eqnarray}
0< N_F^{fd} \leq {3\over 7}(N_F^{adj}-1)& &\cdots 
(\theta_1,\theta_2)=\Bigl({2\over 3}\pi, 0\Bigr)+{\rm permutations},
\nonumber\\
(N_F^{adj}-1)< N_F^{fd} \leq 9(N_F^{adj}-1)& &\cdots 
(\theta_1,\theta_2)=(\pi, 0)+{\rm permutations},
\nonumber\\
9(N_F^{adj}-1) <  N_F^{fd} & &\cdots 
(\theta_1,\theta_2)=\Bigl({2\over 3}\pi, {2\over 3}\pi\Bigr).
\label{hessianb}
\end{eqnarray}
\par
The vacuum configuration at $\beta=\pi$ corresponding to our example 
(\ref{example}) is given by $(\theta_1, \theta_2)
=(\pi, 0)$ and its permutations, for which the 
residual gauge symmetry is $SU(2)\times U(1)$.
Therefore, we observe that the vacuum configuration changes according to the
values of the phase $\beta$. The configuration in Eq. (\ref{example}) starts 
to change as $\beta$ becomes large, keeping $U(1)\times U(1)$ symmetry,
and arrives at $(\theta_1, \theta_2)=(\pi, 0)$ 
at $\beta=\pi$, where $SU(2) \times U(1)$
symmetry is realized\footnote{The gauge symmetry breaking pattern becomes
$SU(3)\rightarrow SU(2)$ for the 
configuration $(\theta_1,\theta_2)=(\pi, 0)$ if we consider 
the nonzero values of $\vev{\phi_{q2}}, \vev{{\bar\phi}_q^{2}}$.}.
\par
What is interesting is that the gauge symmetry breaking pattern
$SU(3)\rightarrow SU(2)\times U(1)$ cannot be 
realized until one considers the softly
broken suppersymmetric QCD with the massless adjoint matter. Actually, as we
have studied in the previous section, the gauge symmetry breaking pattern 
in the softly broken supersymmetric QCD and Yang-Mills 
theory is $SU(N)\rightarrow SU(N)$ and that the 
softly broken supersymmetric gauge theory with only the massless
adjoint matter is $SU(N)\rightarrow U(1)^{N-1}$~\cite{takenagab}. 
This partial gauge symmetry breaking has been pointed out in 
the nonsupersymmetric gauge theory with both the massless adjoint and 
fundamental matter~\cite{intf}.
\par
If we change the number of flavors, the vacuum configuration 
at $\beta=\pi$ also changes. For $(N_F^{adj}, N_F^{fd})=(4, 1)$, the 
vacuum configuration is given by $(\theta_1, \theta_2)=(2\pi/3, \pi/15)$ 
from Eq. (\ref{solsu3}) for the small values of $\beta$, while
at $\beta=\pi$, taking account of Eq. (\ref{hessianb}), it is
given by $(\theta_1, \theta_2)=(2\pi/3, 0)$. The configuration
at $\beta=\pi$ still 
respects $U(1)\times U(1)$ symmetry though the configurations themselves
are different for the two cases.
\par
The above observation implies that if $N_F^{adj}$ increases, then the 
first term in Eq. (\ref{gsqcdeff}) dominates in the effective potential. 
The vacuum configuration
tends to realize the maximal breaking of $SU(3)$. This is consistent with
the result that the dynamics of the nonintegrable phases
for the massless adjoint matter always result in the maximal breaking of 
$SU(N)$, i.e., $U(1)^{N-1}$~\cite{takenagab}. If we, instead, increase 
$N_F^{fd}$ for fixed $N_F^{adj}$, the vacuum configuration tends toward 
the original gauge symmetry. 
This is because the second term in Eq. (\ref{gsqcdeff})  
dominates in the effective potential for a large number of $N_F^{fd}$, and 
the potential has the $SU(N)$ symmetric vacuum as we have studied in the
section $3$.
\par
Let us finally discuss the massless state of 
the adjoint Higgs scalar. To this end, we study
the determinant of the Hessian for the 
configuration $(\theta_1,\theta_2)=(\pi, 0)$,
\begin{equation}
{\rm det}H\bigg|_{\beta=\pi}
=\left(N_F^{fd}-(N_F^{adj}-1)\right)\left(9(N_F^{adj}-1)-N_F^{fd}\right).
\label{hessianc}
\end{equation}
The determinant vanishes for 
the case $N_F^{fd}=N_F^{adj}-1$ or $N_F^{fd}=9(N_F^{adj}-1)$ except for 
the aforementioned ${\cal N}=2$ supersymmetry. The conditions are
satisfied without any fine-tuning of the parameters as long 
as $N_F^{adj}$ and $N_F^{fd}$ are 
discrete numbers. 
In our example, $(N_F^{adj}, N_F^{fd})=(2, 1)$ satisfies 
the former condition. The vanishing determinant implies that the Hessian 
contains the massless mode, which
is nothing but the massless adjoint Higgs scalar 
in our approximation\footnote{This vanishing determinant is modified if we 
consider the nonzero values of the vacuum expectation values for 
the squark fields.}. The massless state of the adjoint Higgs scalar 
has also been pointed out in the nonsupersymmetric 
gauge theories~\cite{intf}.
\par
For comparison to the case of $\beta=\pi$, let us evaluate the second 
derivative of the effective potential (\ref{gsqcdeff}) for the 
small values of $\beta$. 
The vacuum configuration in this case is given by
Eq. (\ref{gsqcdsolb}) and breaks the $SU(N)$ gauge symmetry
to $U(1)^{N-1}$ spontaneously. The second derivative is calculated, using 
Eq. ({\ref{expand}), as
\begin{equation}
{{\del^2V_{GSQCD}}\over{\del\theta_i\del\theta_j}}
={{C_H^{GSQCD}}\over{\pi^2 L^4}}M_{ij},
\qquad 
C_H^{GSQCD}\equiv \beta^2\left(N(N_F^{adj}-1)+N_F^{fd}\right),
\label{massless}
\end{equation}
where $M_{ij}$ is given by Eq. (\ref{matrix}).
The matrix does not have the zero eigenvalue, and the coefficient
$C_H^{GSQCD}$ never vanishes except for the aforementioned ${\cal N}=2$
supersymmetry. Therefore, the adjoint Higgs scalar 
for the small values of $\beta$ is always massive and cannot be massless.
\section{Conclusions and discussion}
We have studied the gauge symmetry breaking patterns through the Hosotani
mechanism (the dynamics of the nonintegrable phases) in supersymmetric 
QCD with $N_F^{fd}$ numbers of the massless fundamental matter and 
its generalized version by introducing $N_F^{adj}$ numbers of the 
massless adjoint matter. The supersymmetry is broken softly by
the Scherk-Schwarz mechanism to give the nonvanishing effective potentials
for the phases.
\par
We have first studied the softly broken supersymmetric Yang-Mills 
theory. The $SU(N)$ gauge symmetry is not 
broken, and there are $N$ vacuum states given 
by Eq. (\ref{symsol}). The $N$ vacua are
physically equivalent, $Z_N$ symmetric and are 
related to each other by the 
gauge transformation with Eq. (\ref{gaugetrf}).  
The fields $A_{\mu}, \lambda$ remain massless
on the vacuum configuration.
\par
By introducing $N_F^{fd}$ sets of the massless 
fundamental matter multiplet, we have obtained 
the softly broken supersymmetric QCD with $N_F^{fd}$ flavors. 
The $SU(N)$ gauge symmetry is not 
broken again in this model, but the vacuum configuration itself 
depends on the number of
color $N$. For $N=$ even, there is a single vacuum state, while
for $N=$ odd, there is a doubly degenerate vacuum state.
The symmetry transformations with 
Eqs. (\ref{inva}) and (\ref{invb}) of the effective potential 
relate the degenerate two vacua. 
The $Z_2$ symmetry is broken by the 
massless matter multiplet belonging to the (anti)fundamental 
representation under $SU(N)$. The adjoint Higgs scalar is always massive
in the two models except for the case of the accidental ${\cal N}=2$
supersymmetry in four dimensions. 
\par
We have also discussed the gauge symmetry breaking patterns in 
the generalized version of supersymmetric 
QCD (supersymmetric QCD with the massless adjoint matter).
We have first studied the case of $SU(2)$ and found 
the vacuum configuration given by Eq. (\ref{su2sol}), which 
breaks the $SU(2)$ gauge symmetry to $U(1)$ spontaneously. 
There is no massless state of the adjoint Higgs scalar in this case.
\par
In order to see how the gauge symmetry is broken through
the Hosotani mechanism for higher rank gauge group, we have 
considered the $SU(3)$ gauge group and chosen the appropriate 
numbers of the flavors as a demonstration.
The vacuum configuration changes according to the values 
of the supersymmetry breaking parameter
$\beta$ by the Scherk-Schwarz mechanism. 
We have explicitly shown that the vacuum 
configurations for small values of $\beta$ and $\beta=\pi$ are given by
the different configurations, which realize the different gauge symmetry 
breaking patterns. It is possible to have the gauge 
symmetry pattern such as $SU(3)\rightarrow
SU(2)\times U(1)$ for the choice 
given by Eq. (\ref{example}) at $\beta=\pi$.
This symmetry breaking pattern is peculiar to the model and is never 
observed in the models studied 
in Ref.~\cite{takenagab} and the section $3$. 
\par
We have investigated the massless state of the adjoint Higgs scalar
by studying the determinant of the Hessian (\ref{hessiana}) for
the small values of $\beta$ and $\beta=\pi$.
We have shown that the massless adjoint Higgs scalar is 
impossible for the small values of $\beta$.
At $\beta=\pi$, however, we have obtained the condition for 
the vanishing determinant of the Hessian without 
any fine-tuning, which implies
the existence of the massless adjoint Higgs scalar in our approximation.
And we have given the explicit example of the parameter choices 
for the massless state. It seems that in order to have the massless 
adjoint Higgs scalar, the partial gauge symmetry breaking 
such as $SU(3)\rightarrow SU(2)\times U(1)$ is necessary. Hence, the 
massless state is a specific feature to the 
generalized version of the softly broken supersymmetric QCD.
\par
We have also discussed the tendency of a gauge symmetry breaking pattern
at $\beta=\pi$ by varying the number of the flavor in 
the generalized supersymmetric QCD. 
If the number of the massless 
adjoint matter $N_F^{adj}$ increases for a 
fixed number of the fundamental matter $N_F^{fd}$, the gauge symmetry 
breaking patterns tend toward the maximal breaking 
of the original gauge symmetry, say, $U(1)\times U(1)$ in our example.
On the other hand, if $N_F^{fd}$ increases for fixed 
$N_F^{adj}$, it tends toward the vacuum configuration respecting the
original gauge symmetry, $(\theta_1,\theta_2)=
({2\over 3}\pi, {2\over 3}\pi)$ in our example. 
\par
It may be interesting to ask what gauge symmetry pattern is realized if we 
consider the higher rank gauge group such as $SU(5)$ in the 
generalized supersymmetric QCD. 
Taking into account the lessons in this paper, one has to 
select carefully the numbers of 
flavors $N_F^{adj}, N_{F}^{fd}$ in order to realize
the partial gauge symmetry breaking such 
as $SU(5)\rightarrow SU(3)\times SU(2)\times U(1)$, which 
may be relevant to
the mechanism of GUT symmetry breaking. We need further 
studies in order to determine the gauge symmetry breaking patterns 
for the higher rank gauge group in the model. 
This will be reported elsewhere.
\par
We have assumed the gauge coupling $g$ is very small and ignored 
the $O(g^2)$ terms in the effective potential.
In this approximation there exists the flat direction
of the potential parametrized by the vacuum expectation values of 
the squark fields, namely $\vev{\phi_q^{adj}}$. 
We have chosen the trivial ``point'' corresponding to
the vanishing vacuum expectation values of them, and we 
have studied the gauge
symmetry breaking patterns through the dynamics of the nonintegrable
phases alone. In order to determine the whole vacuum
structure, one needs to take into account the 
ignored $O(g^2)$ term including the tree-level potential and 
one-loop corrections to the vacuum expectation values of the squark fields. 
\par
We have implicitly assumed the mass term for the squarks, from which we have 
defined the boundary condition of the squark field
associated with the $U(1)_R$ symmetry. We have taken the massless limit 
in order to study the gauge symmetry breaking patterns. It is expected 
that the nonvanishing mass term may also influence the gauge symmetry 
breaking~\cite{hosotanic}. It is 
important and interesting to study the effect of 
the mass term on the Hosotani mechanism.
\vskip 2cm
\begin{center}
{\bf Acknowledgments}
\end{center}  
The author would like to thank the Dublin Institute for Advanced Study
for warm hospitality.
\newpage


\begin{thebibliography}{99}
\bibitem{hosotani} Y. Hosotani, \PLB{126}{309}{83}.
\bibitem{hosotanib} Y. Hosotani, \ANN{190}{233}{89}. 
\bibitem{inta} A. T. Davies and A. McLachlan, \PLB{200}{305}{88}; 
\NPB{317}{237}{89}.
\bibitem{intb} J. E. Hetrick and C. L. Ho, \PRD{40}{4085}{89}.
\bibitem{intc} A. Higuchi and L. Parker, \PRD{37}{2853}{88}. 
\bibitem{intd} C. L. Ho and Y. Hosotani, \NPB{345}{445}{90}.
\bibitem{inte} A. McLachlan, \NPB{338}{188}{90}. 
\bibitem{takenaga} K. Takenaga, \PRD{58}{026004}{98}; {\bf 61}, 
{129902(E)} (2000).
\bibitem{takenagab} K. Takenaga, \PRDM{64}{066001}{01}.
\bibitem{riotto} G. V. Gersdorff, M. Quiros 
and A. Riotto, hep-th/0204041.
\bibitem{ss} J. Scherk and J.H. Schwarz, \PLB{82}{60}{79}. 
\bibitem{fi} P. Fayet, \PLB{159}{121}{85}; \NPB{263}{87}{86}.
\bibitem{takenagac} K. Takenaga, \PLB{425}{114}{98}.
\bibitem{sakamoto} H. Hatanaka, K. Ohnishi, M. Sakamoto and K.Takenaga, 
\PTPM{107}{1191}{02}.
\bibitem{intf} H. Hatanaka, \PTP{102}{407}{99}.
\bibitem{hosotanic} Y. Hosotani, \PLB{129}{193}{83}.
\end{thebibliography}
\end{document}